\documentclass[preprint,
amsmath, amssymb,
aps, pra,
floatfix]{revtex4-2}

\usepackage{graphicx}% Include figure files
\usepackage{dcolumn}% Align table columns on decimal point
\usepackage{bm}% bold math
\usepackage[utf8]{inputenc}
\usepackage[T1]{fontenc}
\usepackage{amsthm}
\usepackage{lipsum}
\usepackage{xcolor}
\begin{document}

\title{Experimental evidence of non-equilibrium phase separation in supercritical fluids}

\author{Juho Lee}
\affiliation{Department of Physics, POSTECH, Pohang, Republic of Korea}
\author{Yeonguk Kim}
\affiliation{Department of Physics, POSTECH, Pohang, Republic of Korea}
\author{Jong Dae Jang}
\affiliation{Neutron Science Division, Korea Atomic Energy Research Institute, Daejeon, Republic of Korea}
\author{Changwoo Do}
\affiliation{Neutron Scattering Division, Oak Ridge National Laboratory, Oak Ridge, TN, USA}
\author{Min Young Ha}
\affiliation{Department of Chemical Engineering, Kyung Hee University, Yongin, Republic of Korea}
\author{Gunsu Yun}
\email{gunsu@postech.ac.kr}
\affiliation{Department of Physics, POSTECH, Pohang, Republic of Korea}
\affiliation{Division of Advanced Nuclear Engineering, POSTECH, Pohang, Republic of Korea}

\begin{abstract}
Supercritical fluids (SCFs) have long been considered homogeneous and structureless, yet recent studies suggest the existence of transient, liquid-like clusters under dynamic processes. In this study, we provide experimental evidence of semi-stable non-equilibrium phase separation in SCFs through opacity measurements and small-angle neutron scattering (SANS). By investigating the thermophysical properties of helium, argon, and krypton during adiabatic expansion, we show that cooling dynamics vary significantly among species, influencing cluster formation. Neutron scattering measurements reveal distinct variations in signal intensity, supporting that the clusters slowly dissolve into the background with a surprisingly long time scale of tens of minutes. Given that SCFs in industrial applications frequently experience dynamic, non-equilibrium conditions rather than in strict thermodynamic equilibrium, our results provide crucial insights with potential implications for advanced material processing, energy systems, and chemical engineering.
\end{abstract}

\maketitle
\section*{Introduction}
Supercritical fluids (SCFs) have long been understood as single-phase systems, distinct from subcritical fluids that exhibit clear liquid and vapor phases. Due to their homogeneous and structureless nature, SCFs are typically regarded as fluids with negligible surface tension and no phase transition. However, recent studies have demonstrated that SCFs can exhibit liquid-like or gas-like properties depending on variations in pressure, temperature, and density \cite{Gorelli2006,Simeoni2010,Brazhkin2012,Brazhkin2013,Banuti2015,Prescher2017,Bryk2017,Pipich2018,Schienbein2018,Maxim2019,Pipich2020}. These variations, although continuous, are associated with phenomena such as sound dispersion, unexpected fluctuations in thermodynamic response functions, and density correlations. The SCF regime at equilibrium can be divided into several sub-regimes separated by the Widom line and Frenkel line. 
The Widom line is an extension of the liquid-gas coexistence curve beyond the critical point, representing the locus of maximum thermodynamic response functions such as heat capacity and compressibility \cite{Gorelli2006, Simeoni2010}. The Frenkel line is defined where the time for a particle to move by its size equals the shortest transverse oscillation period \cite{Brazhkin2012, Brazhkin2013, Bryk2017}. It distinguishes a rigid liquid, where particle motion exhibits solid-like vibrations, from a nonrigid gaslike fluid, where only uncorrelated ballistic motion dominates.
Despite these improved understanding of the thermodynamic characteristics of SCFs, the behavior of SCFs under non-equilibrium conditions remains poorly understood.

Many industrial applications, such as SCF extraction \cite{Taylor1996,Amit2016}, rocket engines \cite{Bellan2020} and high-pressure fuel injection systems, involve highly dynamic non-equilibrium processes, where spatial and temporal variations occur rapidly. In high-pressure fuel injection, for instance, SCFs undergo continuous changes in their thermodynamic and transport properties alongside abrupt transitions like ligament and droplet formation \cite{Bellan2020}. In the transcritical regime, these phenomena create an ambiguous boundary between single-phase and two-phase behavior, underscoring the necessity for systematic investigations into the non-equilibrium properties of SCFs. Beyond industrial applications, non-equilibrium SCF dynamics are also observed in planetary meteorology. For example, the thick atmosphere of Venus consists of SCFs undergoing complex convective and turbulent flows, where rapid thermodynamic changes influence large-scale atmospheric circulation patterns. Understanding these processes requires further exploration of SCF behavior under extreme conditions.

Recently, the non-equilibrium behavior of single-component SCFs has been explored by measuring submicron-sized liquid-like droplets persisting in an argon SCF background \cite{Lee2021}. This study indicates that the extended lifetime of droplets is strongly influenced by the presence of smaller liquid-like particles, known as clusters. These clusters play an essential role in explaining the opacity of high-pressure SCFs and the surprisingly long lifespan of droplets, which can survive up to approximately an hour. Moreover, SCFs with dense, inhomogeneous particles are reported as a promising medium for generating strongly coupled plasmas and extending their lifetimes \cite{Lee2022}. However, while the existence of clusters is theorized, direct experimental evidence of nanometer-scale particles in SCFs has not yet been demonstrated.

\begin{figure*}[t!]
\centering
\includegraphics[width=\linewidth]{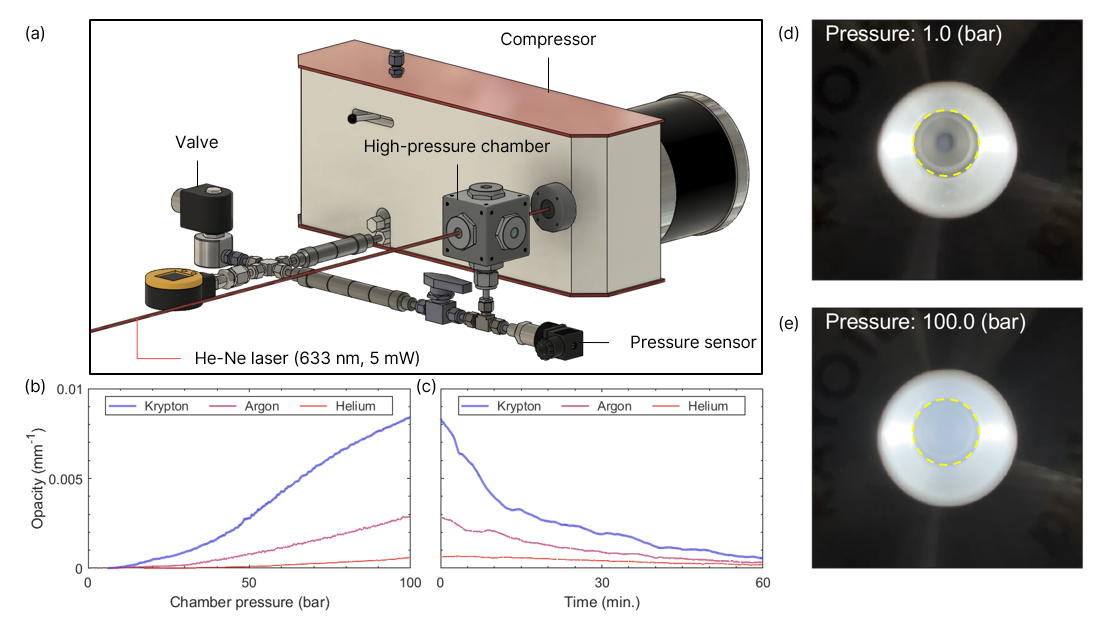}
\caption{\textbf{Experimental apparatus and opacity measurements.} (a) Schematic of the high-pressure chamber system. The He-Ne laser (633 nm, 5 mW) is used to measure the scattered and transmitted signals. (b) Medium opacity within the high-pressure chamber at time zero. Increased opacity indicates the presence of dense particles that scatter the laser. (c) Time dependence of medium opacity at 100 bar. The turbidity inside the chamber gradually decreases, becoming transparent over time. (d, e) Images inside the chamber during the compression process using argon. The pressure corresponds to 1 bar and 100 bar, respectively. The yellow dashed circle indicates the opposite side of the chamber, which is the gas inlet.}
\label{fig:opacity}
\end{figure*}

%In this work, we present the first experimental evidence of clusters in SCFs using small-angle neutron scattering (SANS).
These clusters, which form through adiabatic expansion and cooling, persist as liquid-like fluid packages before evaporating into the gas-like background. These findings challenge the conventional understanding of SCFs as purely homogeneous systems and suggest that non-equilibrium dynamics can drive the emergence of distinct microstructures within SCFs.

The presence of clusters in SCFs may influence mass transport processes in dense planetary atmospheres, such as those of Venus and Jupiter, as well as in engineering applications like power plant cooling systems, pharmaceutical processes, and high-pressure fuel injection. Additionally, these phenomena could play a role in SCF CO$_2$ cleaning techniques for semiconductor fabrication, where transient clustering might impact process efficiency. By uncovering the non-equilibrium phase behavior of SCFs, this study provides a new perspective on the dynamic properties of SCFs.

\begin{figure*}[t!]
\centering
\includegraphics[width=\linewidth]{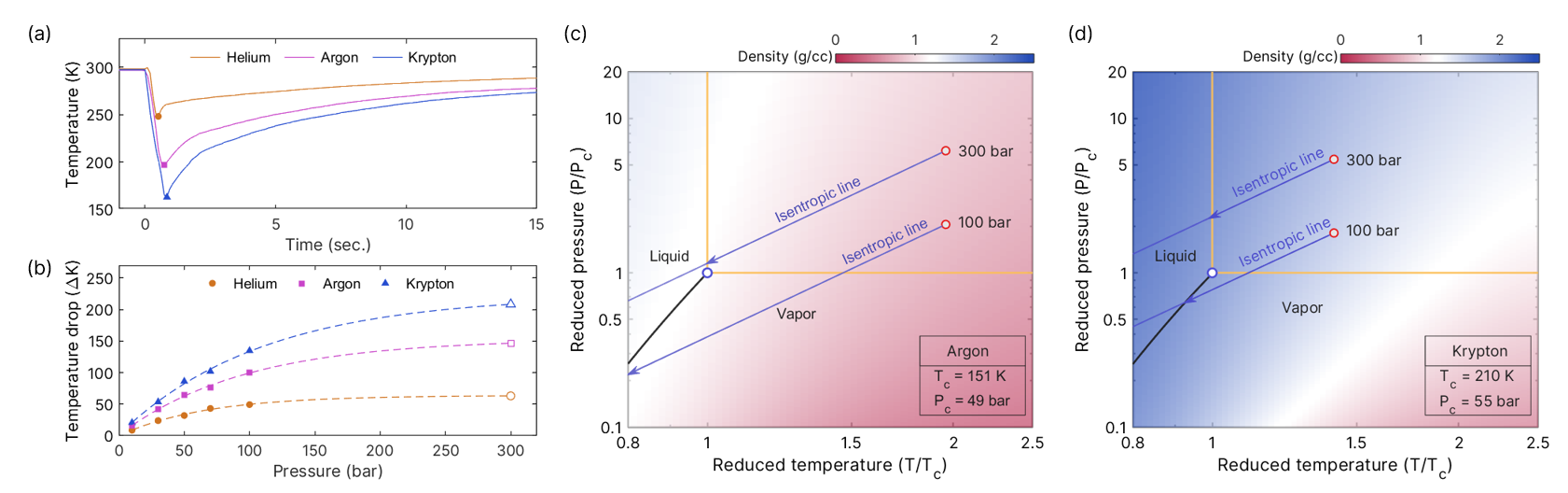}
\caption{\textbf{Formation of liquid-like particles by a temperature drop of expanding fluids.} (a) The experimental measurement for fluid temperature when 100 bar fluid in a high-pressure chamber suddenly expands into a 1 bar environment. The final temperature varies significantly depending on the fluid species. (b) Minimum temperatures for different chamber pressures. A temperature drop of 50 K, 150 K, and 200 K is expected for helium, argon, and krypton, respectively, at the operating pressure of the check valve (300 bar). (c, d) Phase diagram of argon and krypton in equilibrium. Since the compressor operates at a working pressure of 300 bar, the fluid expands into the hose with a relatively lower pressure, leading to cooling. The corresponding isentropic lines are represented by solid lines and arrows. The thermophysical data are obtained from the NIST Chemistry WebBook \cite{NIST2025}.}
\label{fig:temperature}
\end{figure*}

\section*{Experimental apparatus and conditions}

The pressure inside of the high-pressure chamber is increased by the compressor, which consists of the rotating motor and two pistons with check valves. When the motor rotates one cycle, the pistons compress the gas and eject through the check valve opening at the 300 bar. These pistons operate about 4.1 Hz and supply 6.8 cm$^3$ fluid in each cycle. Fig. \ref{fig:opacity}(a) shows the schematic figure of the experimental setup to measure the opacity of SCFs. The high-pressure chamber has five sapphire windows for optical diagnosis, and can hold upto 300 bar pressure without leaking. The opacity is calculated by the ratio between the incident and transmitted laser power (633 nm, He-Ne laser). 

The small angle neutron scattering (SANS) experiments were conducted at the 40-m SANS beamline at the HANARO experimental reactor \cite{Han2013}. The neutron wavelength was 6 $\text{\r{A}}$, and the sample-to-detector distance of 1.16 m was chosen to explore nanometer-scale features in the krypton SCF. The $q$-range spans approximately 0.01 to 0.6 $\text{\r{A}}^{-1}$. The specially designed high-pressure cell for the neutron beam experiment can contain SCF pressure of up to 100 bar. Unlike solid and polymer samples commonly used in neutron scattering experiments, SCFs have lower density and smaller scattering cross-sections, making signal detection more challenging. A sample thickness of 5 mm was implemented to increase the scattering volume. Aluminum windows were chosen to minimize neutron absorption and scattering noise while ensuring safety at high pressure. However, they produce non-negligible scattering at low $q$ . This was not a significant issue, as the clusters of interest primarily appear at higher $q$ values, where the aluminum background is less relevant. The windows have a clear aperture of 16 mm diameter and a thickness of 1 mm.
The windows have a clear aperture of 16 mm diameter and a thickness of 1 mm. The SANS experiment was performed on krypton because it generated the largest number of clusters, as indicated by optical measurements (see Fig. \ref{fig:opacity}). Additionally, while argon has a neutron scattering cross-section of 0.683$\times$10$^{-24}$ cm$^{2}$, krypton has a significantly higher cross-section of 7.68$\times$10$^{-24}$ cm$^{2}$, making it more favorable for scattering signal detection \cite{Sears1992}. To observe the time-dependent evolution of clusters, SANS data was collected every minute. The data has been corrected for detector sensitivity and electrical noise. The 1D scattering intensity was obtained by azimuthally averaging the measured 2D isotropic scattering data. Data points in the low-$q$ and high-$q$ regions, where measurement errors are significant due to detector edge artifacts, were excluded.

\section*{Generation of liquid-like particles by temperature drop}

The opacity of SCFs was measured to demonstrate how their properties change with pressure. The initial He-Ne laser power, $I_0$, and the transmitted power, $I$, were measured to calculate the opacity using the equation:
\begin{equation}
\begin{split}
I = I_0 \exp(-\kappa d),
\end{split}
\end{equation}
where $\kappa$ represents the opacity, and $d$ is the distance the laser travels through the SCF sample medium. Fig. \ref{fig:opacity}(b) shows that the opacity inside the chamber increases as pressure increases. Fig. \ref{fig:opacity}(d) and (e) are the photos looking inside the high-pressure chamber, captured during the compression process. It shows a significant change of opacity depending on pressure. The 100 bar SCF is too opaque to observe the other side of the chamber. The enhanced scattering from high-pressure SCF arises from the generation of dense particles, rather than from the increase in particle density, which is approximately proportional to the pressure. Fig. \ref{fig:opacity}(c) underpins the generation of liquid-like particles during the compression process. When the pressure inside the chamber reaches 100 bar, fogginess is observed, which gradually diminishes over time. Transparency is almost restored after one hour. Thus, the increase in opacity is a temporary phenomenon associated with non-equilibrium phase coexistence, analogous to the evaporation process in the subcritical regime.

The contribution of scattering from the generated clusters to the observed opacity can be estimated using the Rayleigh scattering cross-section. The scattering cross-section of an individual cluster is given by:
\begin{equation}
\begin{split}
\sigma = \frac{2{\pi}^5}{3}\frac{(2r)^6}{\lambda^4}\left(\frac{n^2-1}{n^2+2}\right)^2,
\end{split}
\end{equation}
where r is the radius of particle, $\lambda$ is the wavelength of the incident light, and $n$ is the refractive index ratio between the scattering particle and the surrounding medium. The opacity due to scattering is given by $\kappa_s = n_s \times \sigma$, where $n_s$ is the number density of the particle species.
Assuming that the refractive index of liquid krypton at 150 K is $n \sim$ 1.3 \cite{Sinnock1969}, we can estimate the possible size and density range if their scattering contribution dominates the measured opacity.

Meanwhile, direct imaging of the clusters is not feasible because they are much smaller than the diffraction limit of the 633 nm He-Ne laser. Therefore, to obtain direct evidence of these small particles, we conducted SANS experiments, which will be discussed in the following sections.

Interestingly, the opacity level varies among species. Krypton exhibits a significant increase in opacity, whereas helium remains unchanged from its initial level at 1 bar. This difference arises from variations in their thermophysical properties. The formation of dense liquid-like particles, or clusters, is attributed to localized cooling caused by adiabatic expansion at the compressor outlet. Therefore, the location of the final temperature and pressure point in phase diagrams after the expansion process is essential. Using helium, argon, and krypton gases, we measured the temperature drop for each fluid to quantify the adiabatic expansion process from high pressure to atmospheric pressure. A thermocouple is positioned 10 mm from the outlet of the high-pressure chamber. As shown in Fig. \ref{fig:temperature}(a), the fluid temperature drops sharply when the valve opens. Despite undergoing the same pressure change, the magnitude of the temperature drop is different depending on the fluid species. Fig. \ref{fig:temperature}(b) shows the temperature change as a function of chamber pressure. The check valve inside the compressor opens at 300 bar, which implies that the temperature of krypton is expected to drop by more than 200 K. Such a significant temperature drop would facilitate formation of the liquid-density particles during the expansion process. This is further examined through the reduced phase diagrams of argon and krypton in their equilibrium states, shown in Fig. \ref{fig:temperature}(c,d). Along the isentropic lines depicted in the phase diagrams, the temperature decreases, and upon reaching the liquid phase, a phase transition to the liquid state occurs.
Although the measured temperature drop does not perfectly correspond to that predicted by ideal adiabatic expansion, the minimum temperature experienced by the fluid in the compressor clearly falls within the liquid-phase region of the phase diagram. The transition into the liquid phase occurs more readily in krypton than in argon, which is consistent with the higher opacity observed for krypton in Fig. 1(b). In contrast, helium exhibits a much smaller temperature drop, and its liquid-gas coexistence line is too far from room temperature, preventing it from reaching the liquid phase.

As demonstrated, the expansion process in the compressor induces a significant temperature drop, leading to the formation of liquid-like clusters. In the case of krypton, the temperature can decrease to approximately 150 K when the pressure drops from 100 bar to 1 bar. Nevertheless, the temperature of the stainless steel high-pressure hose and chamber remains at room temperature. The pistons in the compressor, operating at 4.1 Hz, introduce krypton fluid at a rate of approximately 1.1×10$^{-3}$ mol/s. The power required to heat the fluid from 150 K to 300 K is estimated as 
\begin{equation}
\begin{split}
P_{heat} = C_p \cdot N \cdot \Delta T \approx 4.2 \text{W},
\end{split}
\end{equation}
where the average heat capacitance $C_p\sim$25.5 J$\cdot$K$^{-1}$$\cdot$mol$^{-1}$ within the given temperature range under 10 bar.
In contrast, the heat transfer rate of the entire system is significantly higher. Even with a conservative assumption of $\delta$T $\sim$ 10 K, the heat transfer rate is estimated as 
\begin{equation}
\begin{split}
\frac{dQ}{dt} \sim k\frac{2{\pi}rL}{d}{\delta}T \sim 160 \text{W},
\end{split}
\end{equation}
where the thermal conductivity of a stainless steel $k$ $\sim$ 16 W$\cdot$m$^{-1}$K$^{-1}$, demonstrating that the system’s heat transfer rate far exceeds the power required to restore the injected fluid to room temperature. As a result, although the incoming fluids have a low enough temperature for liquid-phase formation, the overall system remains at room temperature.

Observing the temperature decrease caused by repeated adiabatic expansion in the compressor and the changes in opacity inside the high-pressure chamber indicate the presence of a large number of liquid-density particles and their stable persistence within high-pressure SCFs. However, aside from indirect estimations, no direct experimental observation of these clusters has been reported. Here, we utilize a neutron beam to investigate the existence and physical properties of these nanometer-sized clusters.

\section*{SANS results}
\begin{figure*}[t!]
\centering
\includegraphics[width=\linewidth]{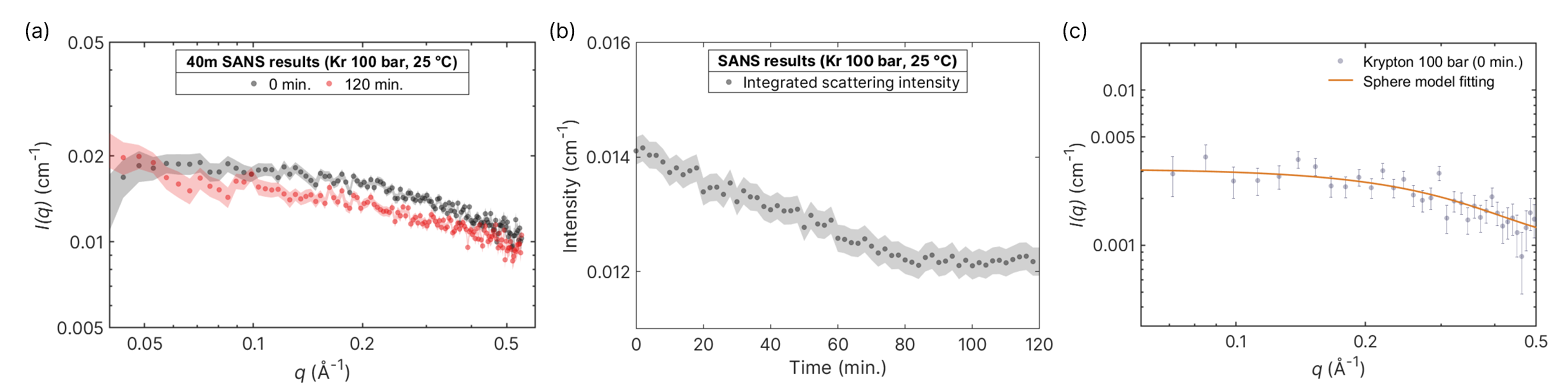}
\caption{\textbf{Neutron scattering signal over time.} (a) After compression, the SANS signal is measured over time. The scattering intensity measured at 2 hours after the compression shows distinct decrease in scattering intensity. (b) Time evolution of the scattering signal. The signal is integrated over the entire q-range of 0.02-0.5 Å$^{-1}$. The scattering intensity gradually decreases over time and saturates around 80 minutes. (c) Spherical model fitting results. The fitting analysis uses the difference between the scattering signals at 0 and 120 minutes. The orange line represents the fitting result, corresponding to scattering from liquid-like particles with a diameter of 1.3 nm.}
\label{fig:SANS}
\end{figure*}

SANS is a powerful tool for analyzing the properties of nanometer-scale particles and structures \cite{Jeffries2021}. Fig. \ref{fig:SANS} shows the neutron scattering results obtained using krypton. Comparing the neutron scattering signals measured immediately after pressurization and 120 minutes later reveals a substantial difference in signal intensity within the given q-range. Fig. \ref{fig:SANS}(b) shows the time evolution of the neutron scattering signal, integrated over the entire $q$-range of 0.02-0.5 Å$^{-1}$. Despite maintaining the same pressure and temperature conditions, the scattering intensity gradually decreases over time and levels off after approximately 80 minutes. The scattering intensity is proportional to the scattering contrast ($\Delta\rho$) and the volume fraction ($\nu$) of the particles:
\begin{equation}
\begin{split}
I(q) \propto \Delta\rho^2  \nu .
\end{split}
\end{equation}
The scattering contrast is given by
\begin{equation}
\begin{split}
\Delta\rho = \rho_l - \rho_g ,
\end{split}
\end{equation}
where $\rho_l$ and $\rho_g$ are the scattering length densities of the liquid-like clusters and the gas-like SCF, respectively. Since the scattering length densities of the clusters and the background medium can be assumed to remain constant, the integrated scattering intensity reflects the volume fraction of the particles. Thus, the gradual decrease in intensity indicates that the clusters responsible for the scattering signal gradually dissolve over an hour.

Interestingly, a measurable slope of the scattering signal is observed even after 120 minutes, indicated by the red points in Fig. \ref{fig:SANS}(a). This signal arises from density fluctuations inherent to the high-pressure krypton SCF. For reference, the correlation length in supercritical CO$_{2}$ and xenon typically ranges from a few angstroms to tens of angstroms, depending on the condition \cite{Sato2008}. To better understand the shape and size of the clusters, we subtracted the scattering contribution from SCF density fluctuations. As shown in Fig. \ref{fig:SANS}(c), the scattering signal is well-fitted by the sphere particle model. The solid line corresponds to a fitting result of a liquid-like krypton particle with an estimated diameter of 1.3 nm, suggesting that it consists of around 30 krypton atoms clustered together.

\begin{figure}[t!]
\centering
\includegraphics[width=85mm]{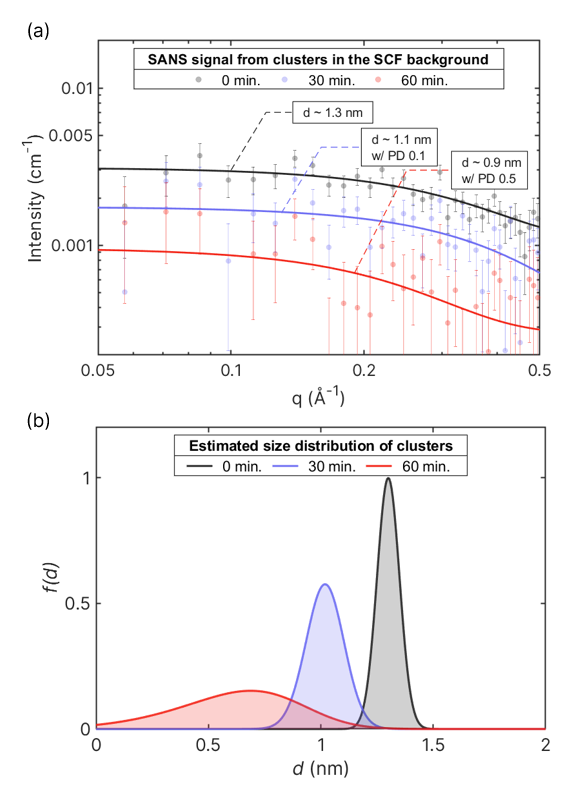}
\caption{\textbf{Time evolution of the scattering signal and cluster size distribution.} (a) Scattering signal at different time points (0 min, 30 min, and 60 min). The data measured at 120 minutes, where the signal no longer decreases, is used as a baseline, and the plotted curves represent the difference from this background. The scattering intensity decreases over time, with increasing error bars, indicating growing uncertainty in the late-time measurements. The fitted curves suggest a reduction in average cluster size over time, accompanied by increasing polydispersity. For the 60-minute data (red curve), the fit includes a polydispersity (PD) value of 0.5, reflecting an increased spread in cluster size distribution. (b) Schematic representation of the expected evolution of the cluster size distribution. Based on the decreasing cluster size, the distributions are estimated to shift toward smaller sizes over time while increasing polydispersity.}
\label{fig:fitting with time}
\end{figure}

\section*{Discussion}
Our results provide the first experimental observation of non-equilibrium phase separation in SCFs. The gradual decrease in scattering intensity over time indicates that liquid-like clusters progressively disappear. It raises a question: how do clusters evaporate into the background SCF? Do they disappear while maintaining their size distribution, or does their size decrease while their number remains constant? Since the fitting of the measured scattering signal represents the average cluster size distribution, analyzing its evolution over time can provide insight into this process.

Fig. \ref{fig:fitting with time} shows the scattering signal from clusters over time with the corresponding fitting results. Similar to Fig. \ref{fig:SANS}(c), this plot shows scattering from the krypton clusters after SCF background subtraction. Initially, the signal is strong but gradually weakens, with increasing error bars. The decrease in volume fraction due to cluster dissipation leads to a reduction in scattering intensity. Notably, after approximately 30 minutes, the overall shape of the graph changes, suggesting an increase in the polydispersity of the cluster size distribution. In liquid evaporation, particle exchange at the surface plays a key role, and smaller particles tend to evaporate more quickly due to their higher surface-to-volume ratio. Consequently, as evaporation progresses, larger particles persist longer while smaller ones disappear, leading to a gradual broadening of the size distribution. As a result, while the early-time signal is well described by a uniform sphere model, the later-time requires consideration of size distribution broadening. This implies that as clusters evaporate, their average size decreases while the overall distribution widens. Fig. \ref{fig:fitting with time}(b) illustrates the estimated evolution of the cluster size distribution over time. The initial distribution at 0 minutes is assumed to be a Gaussian function with a narrow standard deviation. Its mean value is determined based on the fitting results, while the changes in standard deviation and peak height are determined using the total scattering intensity data from Fig. \ref{fig:SANS}(b). The total scattering intensity at 0, 30, and 60 minutes follows a ratio of 1:0.52:0.25, which corresponds to the volume fraction of scattering particles. Since volume fraction depends on the number density of particles and their volumes, the time-dependent size distribution is adjusted so that the product of the integrated distribution (particle number) and the total volume of the particles satisfies this ratio. This ensures a physically consistent representation of how the size distribution broadens and shifts as clusters dissipate over time. However, it remains uncertain whether considering polydispersity is the most appropriate approach, as consideration of polydispersity poses a challenge for signal interpretation \cite{Jeffries2021,Caponetti1993}. This challenge becomes severe in a weak signal, such as the scattering signal at 60 minutes, making precise analysis difficult. Nevertheless, understanding how clusters evolve is essential for capturing the dynamics of their dissipation. In this regard, we hope our study serves as a milestone for further investigations into non-equilibrium phase separation.

\begin{figure}[t!]
\centering
\includegraphics[width=85mm]{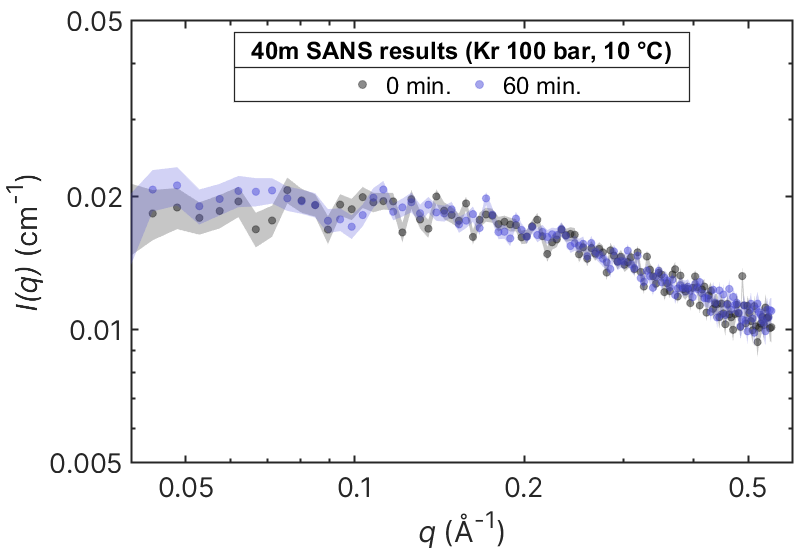}
\caption{\textbf{Time evolution of the neutron scattering signal at a reduced background temperature.} The sample temperature is lowered to 10°C by adjusting the stage temperature of the high-pressure chamber. Unlike the previous results obtained at 25°C, the neutron scattering signal remains stable over time, suggesting a temperature-dependent cluster lifetime.}
\label{fig:SANS low temp}
\end{figure}

The previous SANS measurements are conducted while maintaining the temperature of the background gas-like krypton SCF at room temperature (25°C). If the decaying clusters are analogous to the evaporation process in the subcritical region, this behavior is expected to depend on the temperature of the surrounding medium. To investigate this, we lowered the sample temperature to 10°C by adjusting the stage temperature of the high-pressure chamber. As shown in Fig. \ref{fig:SANS low temp}, the results reveal an intriguing contrast to the previous results at room temperature. When the background SCF is at a lower temperature, the neutron scattering signal remains stable rather than decreasing, indicating that the krypton clusters persist much longer and maintain their opaque state.

\section*{Conclusion}
This study provides experimental evidence for non-equilibrium phase separation in SCFs. While previous research on SCF systems has primarily focused on equilibrium conditions, our findings demonstrate that rapid thermodynamic changes can induce phase separation, producing dense, liquid-like clusters. We observed opacity variations as indirect evidence of clustering, with significant increases in opacity for krypton, while helium remained transparent. To directly confirm cluster formation, we conducted SANS measurements, revealing distinct scattering signals that further support the existence of mesoscopic inhomogeneities within the SCF. An important implication of our findings is that SCFs in industrial applications often experience dynamic pressure and temperature fluctuations rather than remaining in strict equilibrium. The tendency to form dense clusters under such conditions will affect mass transport, solubility, and reactivity. Understanding these non-equilibrium effects expands our knowledge of SCF phase behavior and provides new insights into the design and optimization of SCF technologies.
\begin{figure}[t!]
\centering
\includegraphics[width=85mm]{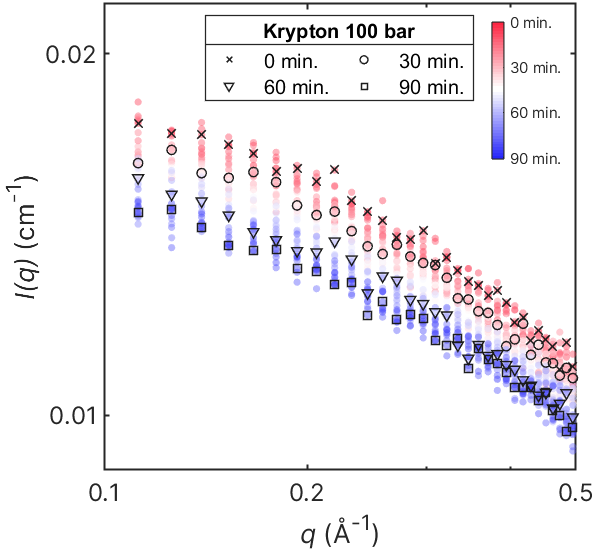}
\caption{\textbf{SANS data over time.} The scattering signals are measured every minute from 0 to 90 minutes and plotted at two-minute intervals after averaging adjacent data points. Distinct markers indicate the measurements at 0, 30, 60, and 90 minutes. As the signal change stabilizes around 80 minutes, the difference between the 60 minutes and 90 minutes data becomes less noticeable.}
\label{fig:appendix}
\end{figure}

%\section*{Data availability}
%All relevant data supporting this study are available from the authors (J.L. and G.Y.) upon reasonable request.

\section*{Acknowledgements}
This work was supported by the National Research Foundation of Korea (NRF) grant funded by the Ministry of Science and ICT (RS-2024-00349684). The internal R$\&$D program at the Korea Atomic Energy Research Institute funded by the Ministry of Science and ICT of Korea (2710007349). A portion of this research used resources at the Spallation Neutron Source, a DOE Office of Science User Facility operated by the Oak Ridge National Laboratory. The beam time was allocated to EQ-SANS on proposal number IPTS-33852.

\section*{Author contributions statement}
G.Y. proposed the original idea and conceived the project. J.L. designed the high-pressure chamber system. J.L., Y.K., and J.D.J. performed the experiments. J.L. carried out the data processing. J.L. and C.D. carried out the data analysis. J.L. wrote the draft of the manuscript. G.Y. supervised the project. All authors contributed to discussions of the results and reviewed the manuscript.

\section*{Appendix}
Fig. \ref{fig:appendix} shows the whole SANS raw data measuring the krypton SCF over time. Data beyond 90 minutes is excluded, as the signal no longer changes and only fluctuates. From 0 minutes (red) to 90 minutes (blue), the gradual weakening of the scattering signal over time can be observed.

\bibliography{references}

\end{document}